\documentclass[a4paper]{jpconf}
\usepackage{graphicx}
\begin{document}
\title{Separable Potentials for (d,p) Reaction Calculations}

\author{Ch. Elster $^{(a)}$,  L. Hlophe$^{(a)}$, V.  Eremenko$^{(a,e)}$
F.M. Nunes$^{(b)}$, I.J. Thompson$^{(c)}$, G. Arbanas$^{(d)}$, J.E. Escher$^{(c)}$}

\address{$^{(a)}$ INPP, Ohio University, Athens, OH, USA}
\address{$^{(b)}$ NSCL, Michigan State University, East Lansing, MI, USA}
\address{$^{(c)}$ Lawrence Livermore National Laboratory, Livermore, CA, USA}
\address{$^{(d)}$ Oak Ridge National Laboratory, Oak Ridge, TN, USA}
\address{$^{(e)}$ SINP, Lomonosov Moscow State University, Moscow, Russia}

\ead{elster@ohio.edu}

\begin{abstract}
An important ingredient for applications of nuclear physics to e.g. astrophysics or
nuclear energy are the cross sections for reactions of neutrons with rare isotopes. Since
direct measurements are often not possible, indirect methods like
$(d,p)$ reactions must be used instead. Those
$(d,p)$ reactions may be viewed as  effective three-body reactions and
described with Faddeev techniques.
An additional challenge posed by $(d,p)$ reactions involving
heavier nuclei is the treatment of the Coulomb force. To avoid numerical complications in
dealing with the screening of the Coulomb force, recently a new approach using the
Coulomb
distorted basis in momentum space was suggested. In order to implement this suggestion,
one needs  to derive a separable representation of neutron- and proton-nucleus
optical potentials and compute their matrix elements in this basis.

\end{abstract}

\section{Introduction}

Nuclear reactions are an important probe to learn about the structure of unstable nuclei.
Due to the short lifetimes
involved, direct measurements are usually not possible. Therefore indirect measurements
using
($d,p$) reactions have been proposed (see e.g.
Refs.~\cite{RevModPhys.84.353,jolie,Kozub:2012ka}).
Deuteron induced reactions are particularly attractive from an experimental perspective,
since deuterated targets are readily available. From a theoretical perspective they are
equally attractive because the scattering problem can be reduced to an effective
three-body
problem~\cite{Nunes:2011cv}. Traditionally deuteron-induced single-neutron transfer
($d,p$) reactions have been used to study the shell structure in stable nuclei, nowadays
experimental techniques are available to apply the same approaches to exotic beams (see
e.g.~\cite{Schmitt:2012bt}).
Deuteron induced $(d,p)$ or $(d,n)$ reactions in inverse kinematics are
also useful to extract neutron or proton capture rates on unstable nuclei of
astrophysical
relevance. Given the many ongoing experimental programs  worldwide using these reactions,
a reliable reaction theory for $(d,p)$ reactions is critical.

One of the most challenging aspects of solving the three-body problem for nuclear reactions
is the repulsive Coulomb interaction.
While the Coulomb interaction for light nuclei is often a small correction to the problem,
this is certainly not the case for intermediate mass and heavy systems. Over the last
decade, many theoretical efforts have focused on advancing the theory
for $(d,p)$ reactions (e.g.  \cite{Mukhamedzhanov:2012qv,Deltuva:2013jna})
and testing existing methods (e.g.  \cite{Deltuva:2007gj,Nunes:2011cv,Upadhyay:2011ta}).
Currently, the most complete implementation
of the theory is provided by the Lisbon group \cite{Deltuva:2009fp}, which solves the
Faddeev equations in the Alt, Grassberger and Sandhas \cite{ags} formulation. The method introduced
in \cite{Deltuva:2009fp} treats the Coulomb interaction with a screening and renormalization
procedure as detailed in  \cite{Deltuva:2005wx,Deltuva:2005cc}. While the current
implementation of the Faddeev-AGS equations with screening is computationally effective for
light systems, as the charge of the nucleus increases technical difficulties arise in the
screening procedure \cite{hites-proc}. Indeed, for most of the new exotic nuclei to be
produced at the Facility of Rare Isotope Beams, the current method is not adequate. Thus one
has to explore solutions to the nuclear reaction three-body problem where the Coulomb
problem is treated without screening.

In Ref.~\cite{Mukhamedzhanov:2012qv}, a three-body theory for $(d,p)$ reactions is derived
with explicit inclusion of target excitations, where no screening of the Coulomb force is
introduced. Therein, the Faddeev-AGS equations are cast in a Coulomb-distorted
partial-wave representation, instead of a plane-wave basis.
This approach assumes the interactions in the two-body subsystems to be separable. 
While in  Ref.~\cite{Mukhamedzhanov:2012qv} the lowest angular momentum in this basis
($l=0$) is derived for a
Yamaguchi-type nuclear interaction is derived as analytic expression,
it is desirable to implement more general form factors, which are modeled after the nuclei
under consideration.

In order to bring the three-body theory laid out in Ref.~\cite{Mukhamedzhanov:2012qv} to
fruition, well defined  preparatory work needs to be successfully carried out.
Any momentum space Faddeev-AGS type calculation needs as input transition matrix elements
in the different two-body subsystems. In the case of ($d,p$) reactions with nuclei these are
the $t$-matrix elements obtained from the neutron-proton, the neutron-nucleus and
proton-nucleus interactions. Since it is essential to use separable interactions when
solving the Faddeev equations in the Coulomb basis, 
those need to be developed not only in the
traditionally employed plane wave basis, but also the basis of Coulomb scattering states.

In this contribution major developments needed to provide reliable input to a
Faddeev-AGS formulation of $(d,p)$ reactions in the Coulomb basis are summarized. Those are
the derivation of separable representations of neutron-nucleus and proton-nucleus optical
potentials. Here it is important that those representations not only describe the cross
sections for elastic scattering accurately, but also allow to represent a wide variety of
nuclei. Furthermore, it should be straightforward to generalize the representation to
account for excitations of the nuclei.

\section{Separable Representation of Nucleon-Nucleus Optical Potentials}

Separable representations of the forces between constituents forming the subsystems in a
Faddeev approach have a long tradition in few-body physics. There is a large body of work on
separable representations of
nucleon-nucleon (NN) interactions (see e.g.
Refs.~\cite{Haidenbauer:1982if,Haidenbauer:1986zza,Berthold:1990zz,Schnizer:1990gf,Entem:2001it})
or  meson-nucleon interactions~\cite{Ueda:1994ur,Gal:2011yp}.
In the context of describing light nuclei like $^6$He~\cite{Ghovanlou:1974zza}
and $^6$Li~\cite{Eskandarian:1992zz} in a three-body approach, separable interactions have
been successfully used.
A separable nucleon-$^{12}$C optical potential was proposed in
Ref.~\cite{MiyagawaK}, consisting of a rank-1 Yamaguchi-type form factor fitted to
the positive energies and a similar term describing the bound states in the
nucleon-$^{12}$C configuration.  However,
systematic work along this line for heavy nuclei,
for which  excellent phenomenological descriptions exist in terms of
Woods-Saxon functions~\cite{Varner:1991zz,Weppner:2009qy,Koning:2003zz,Becchetti:1969zz} has
not been carried out until recently~\cite{Hlophe:2013xca}.
\begin{figure}[h]
\includegraphics[width=18pc]{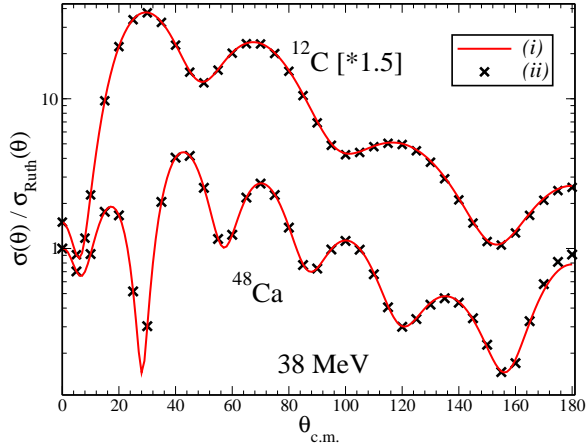}\hspace{2pc}%
\begin{minipage}[b]{14pc}\caption{\label{fig1}
The unpolarized differential cross section for elastic
scattering of protons from $^{12}$C (upper) and $^{48}$Ca (lower) divided by the
Rutherford cross section as function of the c.m. angle calculated for 
E$_{lab}$~=~38~MeV. The $^{12}$C cross section is scaled by a factor 1.5.
The solid lines ($i$) depict the cross
section calculated in momentum space based on the rank-4 separable representation of
the CH89~\cite{Varner:1991zz} phenomenological optical potential, while the cross symbols
($ii$) represent the corresponding coordinate space calculations.
}
\end{minipage}
\end{figure}
The separable representation of two-body interactions suggested by
Ernst-Shakin-Thaler~\cite{Ernst:1973zzb} (EST) is well suited for achieving
this goal. We note that this EST approach
has been successfully employed to represent NN
potentials~\cite{Haidenbauer:1982if,Haidenbauer:1986zza}.
However, the EST scheme derived in Ref.~\cite{Ernst:1973zzb},
though allowing energy dependence of the potentials~\cite{Ernst:1974zzb,Pearce:1987zz},
assumes that they are Hermitian.
Therefore, we generalized the EST approach in Ref.~\cite{Hlophe:2013xca}
in order to be applicable for optical potentials which are complex.
Here it is important that the reaction matrix elements constructed from these separable
potentials satisfy reciprocity relations.

In analogy to the procedure followed in Ref.~\cite{Ernst:1973zzb} we
define a complex separable potential of arbitrary rank in a given partial wave as
\begin{equation}
{\bf U}  = \sum_{i,j} u| f_{l,k_{E_i}} \rangle \langle f_{l,k_{E_i}} |M |
f^*{_l,k_{E_j}}\rangle
\langle f^*_{l,k_{E_j}}|u .
\label{eq:2.1}
\end{equation}
Here $f_{l,k_E}(r)$ is the unique regular radial wave function corresponding to $u$ and
$f^*_{l,k_E}(r)$ is the unique regular radial wavefunction corresponding to $u^*$, where
$u$ is the potential for which the separable representation is constructed. The EST 
scheme guarantees that at the fixed
set of energies ${E_i}$  (support points) the wave functions obtained with the original
potential $u$ and those obtained with the separable representation $U$ are identical. 
The  matrix $M$ is  defined and constrained by
\begin{eqnarray}
\delta_{ik} = \sum_j \langle f_{l,k_{E_i}}|M| f^*_{l,k_{E_j}}\rangle \langle
f^*_{l,k_{E_j}} | u| f_{l,k_{E_k}}\rangle  
 =  \sum_j \langle f^*_{l,k_{E_i}} | u | f_{l,k_{E_j}} \rangle \langle f_{l,k_{E_j}}
|M|f^*_{l,k_{E_k}}\rangle.
\label{eq:2.2}
\end{eqnarray}
The corresponding separable partial wave $t$-matrix must be of the form
\begin{equation}
t(E) =  \sum_{i,j} u| f_{l,k_{E_i}}\rangle \tau_{ij}(E) \langle f^*_{l,k_{E_j}}|u \; ,
\label{eq:2.3}
\end{equation}
with the following  restrictions
\begin{eqnarray}
\delta_{nj} &=&
  \sum_i \langle f^*_{l,k_{E_n}}|u - u g_0(E) u| f_{l,k_{E_i}}\rangle
  \; \tau_{ij}(E),\label{eq:2.4}\\
\delta_{ik} &=&
  \sum_j \tau_{ij}(E) \; \langle f^*_{l,k_{E_j}}|u -u g_0(E) u
  | f_{l,k_{E_k}}\rangle,\label{eq:2.5}
\end{eqnarray}
which are used to obtain the matrix $\tau_{ij}(E)$. In general, optical potentials are
energy dependent. Though the wave functions $f_{l,k_{E_i}}$ carry part of this energy
dependence, the EST scheme needs to be extended to exactly take into account
energy dependent potentials~\cite{Pearce:1987zz,ESTenergy}. 

Extending the EST separable representation to the Coulomb basis involves replacing
   the neutron-nucleus half-shell $t$-matrix in Eq.~(\ref{eq:2.3}) by  Coulomb
distorted scattering states $|f^c_{l,p}\rangle$, which defines the
  Coulomb-distorted separable nuclear $t$-matrix 
\begin{equation}
   \tau^{CN}_l(E) =  \sum_{i,j} u_l^s|f^c_{l,k_{E_i}}\rangle \;\tau^{c}_{i
j}(E)\; \langle f^{c\;,\;*}_{l,\;k_{E_j}}|u_l^s.
 \label{eq:2.6}
\end{equation}
Here $| f^c_{l,k_{E_i}}\rangle$  and $| f^{c\;,\;*}_{l,k_{E_i}}\rangle$ are the  
regular radial Coulomb scattering wave functions,
  corresponding to $u_l^s$  and $(u_l^s)^*$ at energy $E_{i}$.
The constraints are similar to those of Eqs.~(\ref{eq:2.4}) and (\ref{eq:2.5}). The 
Coulomb Green's function is given as 
$g_c (E +i\varepsilon)=(E + i\varepsilon -H_0 - v^C)^{-1}$ 
  with $H_0$ being the free Hamiltonian and $v^C$ the point Coulomb potential. 
It remains to calculate the half-shell
$t$-matrix at the support points in the Coulomb basis. Here we follow the method
suggested in ~\cite{Elster:1993dv} and successfully
  applied in~\cite{Chinn:1991jb}, and note that
  in this case the Coulomb Green's function behaves like a free Green's function. 

For studying the quality of the representation of proton-nucleus optical potentials
  we consider  p+$^{12}$C and  p+$^{48}$Ca elastic scattering and show the
unpolarized differential cross sections divided by the Rutherford cross section
as function of the c.m. angle $\theta_{c.m.}$ in
Fig.~\ref{fig1}.   First, we observe  very good agreement in both cases of the
momentum space
calculations using the separable representation  with the corresponding coordinate
space calculations. Second, we want to point out that we used for the separable
  representation of
  the proton-nucleus partial-wave $t$-matrices the same support points
  as in the neutron-nucleus case. This makes the determination of suitable support
points $E_i$ for a given optical potential and nucleus quite efficient.

\section{Separable Representation of Multi-Channel Optical Potentials}

For representing the NN interaction with a separable interaction, the EST scheme
had to be extended to include channel 
coupling~\cite{Haidenbauer:1982if,Haidenbauer:1986zza} e.g. in the deuteron channel.
For describing $(d,p)$ reactions within a Faddeev-AGS approach, this is a natural
channel coupling to include. However, for $(d,p)$ on nuclei, specifically exotic
nuclei, which are often deformed, one also has to consider possible excitations of
those nuclei. In the formulation of  Ref.~\cite{Mukhamedzhanov:2012qv} the three-body
theory to include excitations
 is layed out. Again, we need to construct separable representations for optical
potentials that include excitations of the nucleus, e.g. rotational 
degrees of freedom. 

To extend the EST scheme to non-Hermitian coupled channel potentials, we define
for a fixed total angular momentum $J$
\begin{eqnarray}
 U^J&=&\sum\limits_{\rho\sigma}\sum\limits_{mn} 
 \left(\sum\limits_{\nu}u\Big|\Psi_{\nu\rho,k_m^\rho}^{(+)J}\mathcal{Y}_{\nu}^{JM}
 \Big\rangle\right) \lambda_{mn}^{\rho \sigma}
\left(\sum\limits_{\nu}\Big\langle\Psi_{\nu\sigma,k_n^\sigma}^{(-)J}
\mathcal{Y}_{\nu}^{JM}\Big|u\right)\cr
    &\equiv&\sum\limits_{\rho\sigma}\sum\limits_{mn}
T(E_m)\Big|\mathcal{Y}_{\rho}^{JM}\phi_{l_\rho k_m}\Big\rangle \lambda_{mn}^{\rho\sigma} 
\Big\langle\phi_{l_\sigma
k_n^\sigma}\mathcal{Y}_{\sigma}^{JM}\Big|T(E_n),
\label{eq:2.7}
\end{eqnarray}
where $\langle\Psi_{\nu\sigma,k_n^\sigma}^{(+)J}|$ are the solutions of the
coupled-channel Lippmann-Schwinger equation corresponding to $u$ and
$\langle\Psi_{\nu\sigma,k_n^\sigma}^{(-)J}$) those corresponding to $u^\dagger$
with incoming boundary conditions. The indices $\rho$ and $\sigma$ indicate the
coupling between channels, while $m$ and $n$ refer to the rank of the separable
representation. 
The constraints necessary to satisfy the EST conditions become
\begin{eqnarray}
 \delta_{mi}\delta_{\rho\gamma}&=&\sum\limits_{\sigma n}\sum\limits_{\nu}
 \lambda_{mn}^{\rho\sigma} 
  \Big\langle\phi_{l_\sigma k_n^\sigma}\mathcal{Y}_{\sigma}^{JM}\Big|T(E_n)\Big|
  \mathcal{Y}_{\gamma}^{JM}\Psi_{\nu\gamma,k_i^\gamma}^{(+)J}\Big\rangle, \cr
   &=&\sum\limits_{\sigma n}\sum\limits_{\nu} 
   \Big\langle\Psi_{\nu \gamma,k_i^\gamma}^{(-)J}\mathcal{Y}_{\nu}^{JM}\Big|T(E_n)\Big|
\mathcal{Y}_{\sigma}^{JM} \phi_{l_\sigma k_n^\sigma}\Big\rangle\lambda_{nm}^{\sigma\rho}.
\label{eq:2.8}
\end{eqnarray}
The multi-channel separable $t$-matrix then takes the form
\begin{eqnarray}
  t_{\alpha\beta}^J(k',k;E)&=&\sum\limits_{\rho\sigma}\sum\limits_{mn}
T^{J}_{\alpha\rho  }(k',k_m^\rho;E_m)\tau_{mn}^{\rho\sigma}(E) 
{T}^{J}_{\sigma \beta }(k,k_n^\sigma;E_n),\cr
    &=&\sum\limits_{ab} T^{J}_{\alpha a}(k',k_a;E_a)\tau_{ab}(E)
     {T}^{J}_{b \beta}(k,k_b;E_b),
\label{eq:2.9}
\end{eqnarray}
where the quantities  $T^{J}_{\alpha \rho}(k',k_a;E_a)$ represent the half-shell
transition matrix elements in channels $\alpha$ and $\rho$. The coupling matrix 
$\tau_{mn}^{\rho\sigma}(E)$ depends now on the rank, indicated by the indices $m$ and $n$ as
well as on the channels indicated by the superscripts. Simplyfing the quadrupel sum of the first
row in Eq.~(\ref{eq:2.9}) over rank as well as channel indices to a double sum over leads to
the second row. Explicitly, the coupling matrix $\tau (E)$ is calculated as
\begin{eqnarray}
   [\tau(E)]_{ab}^{-1}& = &\Big\langle\phi_{l_a k_a} 
\mathcal{Y}_{\alpha_a}^{  JM}\Big|T(E_a)\Big|\mathcal{Y}_{\alpha_b}^{JM} 
 \phi_{l_b k_b}\Big\rangle \\
&+& \sum\limits_{\beta}\int\limits_0^\infty
dp\;p^2\;\Big\langle\phi_{l_ak_a}\mathcal{Y}_{\alpha_a}^{JM}\Big|T(E_a)\Big|
   \mathcal{Y}_{\beta}^{JM}\phi_{l_\beta p}\Big\rangle
 G_{\beta}(E_b)\Big\langle\phi_{l_\beta p}\mathcal{Y}_{\alpha_a}^{JM}|T(E_b)\Big|
   \mathcal{Y}_{\alpha_b}^{JM}\phi_{l_b k_b}\Big\rangle \cr
&-&\sum\limits_{\beta}\int\limits_0^\infty p^2dp\;
   \Big\langle\phi_{l_a k_a}\mathcal{Y}_{\alpha_a}^{JM}\Big|T(E_a)\Big|
\mathcal{Y}_{\beta}  ^{JM}\phi_{l_\beta p}\rangle G_{\beta}(E)
   \Big\langle\phi_{l_\beta
p}\mathcal{Y}_{\beta}^{JM}|T(E_b)\Big|\mathcal{Y}_{\alpha_b}^{  JM}\phi_{l_b
k_b}\Big\rangle. \nonumber
  \label{eq:2.10}
\end{eqnarray}
Here $G_{\beta}(E)=(E-\varepsilon_\beta-E_{p}^\beta+i\epsilon)^{-1}$ is the channel
Green's function with $E_{p}^\beta=p^2/  2\mu_\beta$, with $\mu_\beta$ being the reduced
mass in the channel $\beta$.

\vspace{1mm}

\begin{figure}[h]
\includegraphics[width=20pc]{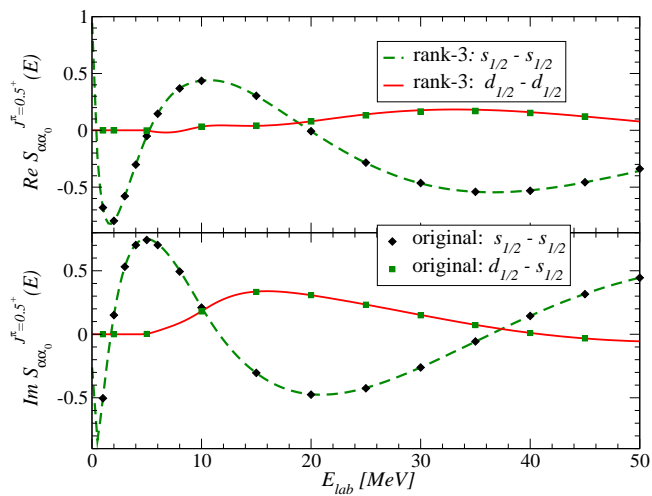}\hspace{2pc}%
\begin{minipage}[b]{14pc}\caption{\label{fig2}
The separable representation of rank 3 for the s-d multichannel S-matrix 
 $S_{\alpha\alpha_0  }^J(E)$
for $n$+$^{12}$C scattering as function of the laboratory kinetic energy for
$J^\pi=\frac{1}{2}^+$. The $0^+\otimes s_{1/2}-0^+\otimes s_{1/2}$  and $2^+\otimes
d_{1/2}-0^+\otimes s_{1/2}$
    $S$-matrix elements are indicated  by the (green) dashed and (red) solid lines.
 The  EST support
points are given at 6, 20, and 40~MeV. The $S$-matrix elements 
obtained using the original potential are shown for comparison.
}
\end{minipage}
\end{figure}

As example for constructing a separable optical potential including rotational
excitations we use the potential from Ref.~\cite{Olsson:1989npa} which describes
elastic and inelastic scattering of neutrons from $^{12}$C in the energy range 16.5 to
22~MeV. This potential includes the excitation  $I^\pi=2^+$ and $I^\pi=4^+$ at 4.44 MeV 
and 14.08~MeV in $^{12}$C, and takes into account quadrupole and octupole deformations.
In Fig.~\ref{fig2} we focus on the on-shell properties of the potential up to 50~MeV 
laboratory kinetic energy. The EST support points are chosen to be at 6, 20, and
45~MeV.  In  Fig.~\ref{fig2} the multi-channel $S$-matrix elements are depicted for
calculations based on the original potential and on its separable
representation. The calculations indicate that the extension of the EST
formulation which includes excitations of the nucleus represents the original optical
potential very well, and thus those potentials can serve as input for $(d,p)$
reaction calculations. 

\section{Summary and Outlook}
In a series of steps we developed the input that will serve as a  basis for
Faddeev-AGS three-body calculations of
  $(d,p)$ reactions, which will not rely on the screening of the
  Coulomb force. To achieve this, Ref.~\cite{Mukhamedzhanov:2012qv} formulated the
  Faddeev-AGS equations in the Coulomb basis using separable interactions in the
two-body subsystems.
For this ambitious program to have a chance of being successful, the
  interactions in the two-body subsystems, namely the NN and the neutron- and
  proton-nucleus systems, need to developed so that they separately describe the
observables 
  of the subsystems. While for the NN interaction separable representations are
available,
  this is was not the case for the optical potentials describing the nucleon-nucleus
  interactions. Furthermore, those interactions in the subsystems need to be
available in the Coulomb basis.

We developed separable representations of phenomenological optical
  potentials of Woods-Saxon type for neutrons and protons. First we concentrated on
  neutron-nucleus optical potentials and generalized the Ernst-Shakin-Thaler (EST)
  scheme~\cite{Ernst:1973zzb} so that it can be applied to complex
potentials~\cite{Hlophe:2013xca}. In order to consider proton-nucleus optical
potentials,
  we further extended the EST scheme so that it can be applied to the scattering of
charged particles with a repulsive Coulomb force~\cite{Hlophe:2014soa}.
  While the extension of the EST scheme to charged particles led to a separable
  proton-nucleus $t$-matrix in the Coulomb basis, we had to develop  methods to
  reliably compute Coulomb distorted neutron-nucleus $t$-matrix
elements~\cite{upadhyay:2014}. Here we also show explicitly that those  calculations
  can be carried out numerically very accurately by calculating them within two
independent schemes.
We also showed that the scheme can be further extended to take channel coupling into
account. 

 Our results demonstrate, that our  separable representations
 reproduce standard coordinate space
calculations of neutron and proton scattering cross sections very well, and that we
are able to
  accurately compute the integrals leading to the Coulomb distorted form factors. Now
that these challenging form factors have been obtained, they can be introduced into the
  Faddeev-AGS equations to solve the three-body problem without resorting to
screening. Our
  expectation is that solutions to the Faddeev-AGS equations written in the
  Coulomb-distorted basis can be obtained for a large variety of  $n+p+A$ systems,
without a limitation on the charge of the target.
  From those solutions,  observables for $(d,p)$
transfer reactions should be readily calculated. Work along these lines is in
  progress.

\section*{Acknowledgments}
This material is based on work  in part supported
by the U.~S.  Department of Energy, Office of Science of Nuclear Physics
 under  program No. DE-SC0004084 and  DE-SC0004087 (TORUS Collaboration), under
contracts   DE-FG52-08NA28552  with
 Michigan State University, DE-FG02-93ER40756 with Ohio University;  by  Lawrence
Livermore
  National Laboratory under Contract DE-AC52-07NA27344 and the U.T. Battelle LLC
Contract   DE-AC0500OR22725.
  F.M. Nunes  acknowledges support from the National Science Foundation
  under grant PHY-0800026. This research used resources of
  the National Energy Research Scientific Computing Center, which is supported by the
Office of
  Science of the U.S. Department of Energy under Contract No. DE-AC02-05CH11231.

\section*{References}
\bibliographystyle{iopart-num}
  
\bibliography{coulomb}

\end{document}